# Orbital Torque in Rare-Earth Transition-Metal Ferrimagnets

Shilei Ding*, Min-Gu Kang, William Legrand, Pietro Gambardella*
*Department of Materials, ETH Zurich, 8093 Zurich, Switzerland*

**Orbital currents have recently emerged as a promising tool to achieve electrical control of the magnetization in thin-film ferromagnets. Efficient orbital-to-spin conversion is required in order to torque the magnetization. Here we show that the injection of an orbital current in a ferrimagnetic $Gd_yCo_{100-y}$ alloy generates strong orbital torques whose sign and magnitude can be tuned by changing the Gd content and temperature. The effective spin-orbital Hall angle reaches up to -0.25 in a $Gd_yCo_{100-y}/CuO_x$ bilayer compared to +0.03 in $Co/CuO_x$ and +0.13 in $Gd_yCo_{100-y}/Pt$. This behavior is attributed to the local orbital-to-spin conversion taking place at the Gd sites, which is about five times stronger and of the opposite sign relative to Co. Furthermore, we observe a manyfold increase in the net orbital torque at low temperature, which we attribute to the improved conversion efficiency following the magnetic ordering of the Gd and Co sublattices.**

The orbital magnetic moment of metals is usually quenched by electron delocalization and crystal field effects. However, an electric field can induce a net flow of nonequilibrium orbital momenta in electric conductors in which the orbital character of the electronic bands varies in momentum space [1-6]. Electric-field-induced orbital currents and polarization are exemplified by the orbital Hall effect (OHE) [2-7] and orbital Rashba-Edelstein effect (OREE) [8-11], which occur in bulk conductors and interfaces with broken inversion symmetry, respectively. The orbital momenta generated by an electric current in a nonmagnetic layer can thus diffuse into an adjacent magnetic layer and exert a torque on the local magnetization (Fig. 1) [12-24], analogously to the spin-orbit torques based on the spin Hall and Rashba-Edelstein effects [25]. The concept of orbital torque opens unprecedented opportunities in spintronics, because it significantly expands the type and number of materials that can mediate the transfer of angular momentum from a charge current to a spin system. Theoretical calculations [2,6,26] and experiments [23,27] show that giant orbital Hall conductivities are present in the $3d$ metals, which are comparable to the spin Hall conductivity of the $5d$ heavy metals [25,28] and result in large orbital torque efficiency [15,20-24], current-induced magnetization switching [20,27], and unconventional magnetoresistance [29-31] in $3d$ metal/ferromagnet bilayers.

The conversion of orbital to spin momenta requires spin-orbit coupling, which is key to the generation of orbital torques. Recent experiments have shown that efficient conversion is achieved in ferromagnets such as Ni with spin-orbit split bands near the Fermi level [19,20] as well as by inserting a $5d$ [32,33,34] or $4f$ [23] metal spacer in between the nonmagnetic $3d$ metal and the ferromagnet. Several studies have investigated the orbital torque efficiency in transition-metal (TM) ferromagnets, showing that the magnitude of the orbital torque changes for different ferromagnets even if the same $3d$ metal is used as a source of the orbital moment [19,20,23]. Despite their fundamental and applied interest, orbital torques in ferrimagnetic metals have not been studied thus far. In particular, rare-earth (RE)-TM amorphous ferrimagnets such as GdCo and GdFeCo are prime candidates for spintronic applications because of their tunable magnetization and fast magnetization dynamics, which result from their antiferromagnetically-coupled sublattices [35-37]. By adjusting the composition of the alloy or by changing the temperature, the magnetization can be tuned from RE- to TM-dominated, reaching the two compensation points where either the net magnetization or the net angular momentum cancels out [35], inducing a divergence of the current-induced spin-orbit torques [38-41] and domain wall velocity [42,43], respectively. The magnetotransport properties of RE-TM alloys, such as the anomalous Hall effect (AHE) [38] and spin-orbit torques [44], are dominated by the TM sublattice. The orbital-to-spin conversion efficiency of ferrimagnetic metals, however, is presently unknown. Remarkably, the large spin-orbit coupling of the RE atoms may offer novel prospects for tuning the orbital-to-spin conversion by changing the ratio of TM to RE atoms.

In this work, we demonstrate the generation of strong orbital torques in the model RE-TM ferrimagnet $Gd_yCo_{100-y}$ coupled to a partially oxidized $CuO_x$ layer, which serves as the source of nonequilibrium orbital currents induced by the OREE [10,14-17,21,24,29-33]. We show that the presence of the RE atoms in the magnetic layer enhances the orbital-to-spin conversion owing to the large spin-orbit coupling of the $5d$ states of Gd. The sign of the orbital torque changes from positive in $Co/CuO_x$ to negative in $Gd_yCo_{100-y}/CuO_x$ due to the opposite spin-orbit coupling of the Co and Gd atoms. The ratio of the angular momentum current to the applied electric field, namely the effective spin-orbital Hall conductivity, varies linearly with Gd content from $-3.9 \pm 0.2 \times 10^4 \, \Omega^{-1}m^{-1}$ ($y = 20\%$) to $-7.8 \pm 0.9 \times 10^4 \, \Omega^{-1}m^{-1}$ ($y = 35\%$) at room temperature. Finally, we find that the magnitude of the spin-orbital Hall conductivity increases by more than a factor of 10 below 100 K, reflecting the strong increase of the OREE in $CuO_x$ and the increase of magnetic order in the GdCo layer. Overall, our work shows that RE-TM ferrimagnets allow for efficient orbital-to-spin conversion and large orbital torques that can be tuned by changing the stoichiometry of the alloy and the degree of ferrimagnetic order.

The $Gd_yCo_{100-y}(10)/CuO_x(6)$ bilayers (thicknesses in nanometers) were grown by dc magnetron sputtering on $Si/SiO_2$ substrates with a $Si_3N_4(6)$ buffer layer. The thickness of the $CuO_x$ layer is given before natural oxidization. The proportion of Gd in the ferrimagnetic alloy varies from 10% to 35% by tuning the deposition rate of Co and Gd in a confocal sputtering configuration [45]. We indicate by $CuO_x$ a layer of Cu naturally oxidized in air. Previous studies of $CuO_x$ have shown that the oxygen gradient in naturally oxidized Cu is critical for the generation of orbital polarization by the OREE [10,14-17,21, 24-33]. A single layer of $CuO_x(3)$ is electrically insulating whereas $CuO_x(6)$ has a resistivity of $1.7 \times 10^{-7}$ $\Omega$m, thereby allowing a current to flow in between the oxide region and the magnetic layer. The root mean square surface roughness of the samples was less than 0.3 nm, as measured by atomic force microscopy [45]. Hall bar devices with a width of 10 μm, a length of 10 μm and an aspect ratio of ~1 were patterned by photolithography and lift-off [Fig. 2(a)]. Harmonic Hall voltage measurements were carried out to measure the orbital torques using an alternate excitation current with a frequency of ~10 Hz [45,46]. All the $Gd_yCo_{100-y}$ samples investigated in this work have in-plane magnetization. Their magnetic properties are reported in Ref. [45].

Figure 2(b) shows the effective magnetic field $\boldsymbol{B}_{DL} \propto \boldsymbol{M} \times \boldsymbol{L}$ corresponding to the damping-like orbital torque measured by the harmonic Hall voltage method in $Co(2)/CuO_x(7)$,

Gd$_{20}$Co$_{80}$(10)/CuO$_x$(6), and a control sample of Gd$_{20}$Co$_{80}$(10) capped by Si$_3$N$_4$(6). Here **M** is the sample magnetization and **L** the orbital momentum injected into the magnetic layer. The harmonic Hall method detects the current-induced oscillations of the magnetization that cause a second harmonic component in the Hall resistance due to the AHE and planar Hall effect, as described in detail in Refs. [45,46]. We find that $B_{DL}$ increases proportionally to the applied electric field (or current) as expected. However, $B_{DL}$ changes from positive in Co(2)/CuO$_x$(7) to negative in Gd$_{20}$Co$_{80}$(10)/CuO$_x$(6). This striking behavior is not compatible with the injection of a spin current from CuO$_x$ because, in such a case, $B_{DL}$ would have the same sign in both samples. Instead, the inversion of $B_{DL}$ is a signature of the injection of orbital momenta from CuO$_x$ into Co and Gd$_y$Co$_{100-y}$, as the ensuing orbital torques depend on the sign of the spin-orbit coupling in the magnetic layer [13,19,20,23].

Fitting the electric field dependence of $B_{DL}$ shown in Fig. 2(b) allows us to obtain the orbital torque efficiency per unit applied electric field, $\xi_{DL}^E = \frac{2e}{\hbar} M_s t_{GdCo} B_{DL}/E$. Here, $e$ is the electronic charge, $\hbar$ the reduced Planck's constant, $M_s t_{GdCo}$ the product of the saturation magnetization and thickness of the Gd$_y$Co$_{100-y}$ layer, and $\xi_{DL}^E$ represents the effective spin-orbital conductivity of the bilayer in units of $\Omega^{-1}m^{-1}$. The electric field is given by $E = I \frac{R_{xx}^{1\omega}}{l}$, where $I$ is the current, $R_{xx}^{1\omega}$ the first harmonic longitudinal resistance and $l$ the length of the Hall bar. We thus find $\xi_{DL}^E = (7.2 \pm 0.8) \times 10^4 \, \Omega^{-1}m^{-1}$ in Co(2)/CuO$_x$(7) and $\xi_{DL}^E = (-3.9 \pm 0.2) \times 10^4 \, \Omega^{-1}m^{-1}$ in Gd$_{20}$Co$_{80}$(10)/CuO$_x$(6). Remarkably, $\xi_{DL}^E$ of Gd$_{20}$Co$_{80}$(10)/CuO$_x$(6) is also opposite to that of Gd$_{20}$Co$_{80}$(10)/Pt(5) [Fig.3], which is attributed to the spin current injected from Pt. Measurements of a control sample without CuO$_x$ show that $\xi_{DL}^E = (0.03 \pm 0.04) \times 10^4 \, \Omega^{-1}m^{-1}$ in Gd$_{20}$Co$_{80}$(10)/Si$_3$N$_4$(6), excluding the possibility of a significant self-torque in the Gd$_y$Co$_{100-y}$ layer [45,53]. As we demonstrate below, the competition between the orbital-to-spin conversion efficiency of Co and Gd determines the overall sign and magnitude of the orbital torque in Gd$_y$Co$_{100-y}$.

Figure 3(a) shows the saturation magnetization $M_s = |M_{Co} - M_{Gd}|$ and anomalous Hall resistance ($R_{AHE}$) of Gd$_y$Co$_{100-y}$(10)/CuO$_x$(6) in samples with different amounts of Gd. Upon increasing the Gd proportion in Gd$_y$Co$_{100-y}$, the ferrimagnetism changes from Co-dominated to Gd-dominated, and $R_{AHE}$ becomes negative as the Co magnetization ($M_{Co}$) aligns antiparallel to the external field. Both anomalous Hall and magnetic measurements indicate that the magnetic compensation point is located around $y = 27\%$ in this alloy series. Figure 3(b) presents the orbital torque efficiency $\xi_{DL}^E$ as a function of Gd content. The data show that compensation between the orbital torque promoted by Co and Gd is achieved around $y = 15\%$. Furthermore, the change in the orbital torque efficiency with different proportions of Gd and Co is approximately linear in $y$ and can be fit by

$$\xi_{DL}^E = T_{int} \frac{2e}{\hbar} \sigma_O^{CuOx} \left[ \frac{y}{100} \eta_{L \cdot S}^{Gd} + \frac{100-y}{100} \eta_{L \cdot S}^{Co} \right], \tag{1}$$

where $T_{int}$ is the orbital transmission coefficient of the Gd$_y$Co$_{100-y}$/CuO$_x$ interface, $\sigma_O^{CuOx}$ is the orbital conductivity of CuO$_x$ and $\eta_{L \cdot S}^{Gd} < 0$ and $\eta_{L \cdot S}^{Co} > 0$ represents the orbital-to-spin conversion efficiency of Gd and Co, respectively. In this scenario, the negative orbital torque of Gd$_y$Co$_{100-y}$/CuO$_x$ originates from the negative and large spin-orbit coupling of the 5$d$ states of Gd and their hybridization with the 3*d* states of Co [53-56]. The competition between the spin-orbit coupling of Co and Gd atoms thus results in a tunable orbital-to-spin conversion. This composition dependence is a distinctive feature of the orbital torque relative to the spin torque. In RE-TM ferrimagnets coupled to a source of spin current such as Pt, the spin torque efficiency does not change significantly with the RE-TM ratio [squares in Fig. 3(b) and Refs. 38,44].

Given that the distribution of RE and TM atoms in GdCo is highly inhomogeneous, as typical of amorphous RE-TM alloys [44,57,58], the magnetic properties of GdCo, such as $M_s$, magnetic anisotropy, and Curie temperature, do not scale linearly with Gd content [59-61]. Moreover, the average magnetic moment of the TM atoms depends on the number of TM-atom nearest neighbors, as described by the so-called environment model [59-61]. With these premises, the approximate linear scaling of $\xi_{DL}^E$ vs $y$ indicates that the enhanced orbital-to-spin conversion efficiency introduced by Gd is mostly a single-ion property of the RE atoms. Accordingly, $\eta_{L \cdot S}^{Gd}$ depends on the average concentration of Gd rather than on the inhomogeneous distribution of Gd-rich and Co-rich phases in the alloy. From the compensation of $\xi_{DL}^E$ at $y = 15\%$ we obtain $|\eta_{L \cdot S}^{Gd}| \gtrsim 5|\eta_{L \cdot S}^{Co}|$, which is consistent with $\xi_{DL}^E$ being dominated by the concentration of Gd atoms. The ratio $\eta_{L \cdot S}^{Gd}/\eta_{L \cdot S}^{Co}$ is consistent with the larger spin-orbit coupling of the Gd 5d states (~ -250 meV) [54] relative to that of the Co 3d states (~60 meV). The positive value of $\eta_{L \cdot S}^{Co}$ agrees with other studies of the orbital torque in Co/CuO$_x$ [30,62]. Torque measurements do not allow for determining $\eta_{L \cdot S}^{Co}$ and $\eta_{L \cdot S}^{Gd}$ individually, even in elemental Co and Gd samples, because the parameters $T_{int}$ and $\sigma_O^{CuOx}$ in Eq. (1) cannot be separately measured. An estimate of $T_{int}\eta_{L \cdot S}^{Co} \approx 3.6 \%$ in Co/CuO$_x$ can be obtained by taking $\sigma_O^{CuOx} = 1 \times 10^6 \left(\frac{\hbar}{e}\right)\Omega^{-1}m^{-1}$ from Ref. [10] and $\xi_{DL}^E = T_{int} \frac{2e}{\hbar}\sigma_O^{CuOx}\eta_{L \cdot S}^{Co} = (7.2 \pm 0.7) \times 10^4 \Omega^{-1}m^{-1}$ from our experiments. This estimate agrees with theory [19] and provides a lower boundary for $\eta_{L \cdot S}^{Co}$, showing that there is ample room for increasing the orbital torque by improving $\eta_{L \cdot S}$ and $T_{int}$ by material engineering.

Deviations from the linear trend exemplified by Eq. (1) can be ascribed to environment effects related to the change of TM magnetic moment and the TM-TM and TM-RE exchange coupling constants [59-61], as well as to unintentional variations of the alloy structure and composition in samples fabricated in different batches using variable sputtering power [45]. The extrapolation of the linear fit to $y = 0$ further shows that $\xi_{DL}^E$ predicted by Eq. (1) becomes positive but remains below the efficiency of Co(2)/CuO$_x$(7). A possible explanation of the difference between the extrapolated and measured values at $y = 0$ is the different thickness of the Co layer in the Co/CuO$_x$ and Gd$_y$Co$_{100-y}$/CuO$_x$ samples, which affects the total amount of angular momentum converted and absorbed from CuO$_x$. We also note that the orbital transmission coefficient $T_{int}$ can vary with $y$ and thus introduce further deviations from the linear scaling assumed in our fit. There are presently no accurate models available to describe the dependence of $T_{int}$ on interface crystallinity and disorder [12], which calls for further investigations of this parameter.

To better understand the factors that affect the orbital torque in ferrimagnets we measured the temperature dependence of $\xi_{DL}^E$. We focus first on Gd$_{23}$Co$_{77}$(10)/CuO$_x$(6), whose magnetic compensation temperature is around 120 K, as shown in Fig. 4(a). Figure 4(b) compares the temperature dependence of $\xi_{DL}^E$ in $Gd_yCo_{100-y}(10)/CuO_x(6)$ with that of $Co(2)/CuO_x(7)$, $Gd_{23}Co_{77}(10)/Pt(5)$. Whereas the spin torque efficiency per applied electric field in $Gd_{23}Co_{77}(10)/Pt(5)$ only increases by 25% from 300 to 20 K, the orbital torque efficiencies in $Co(2)/CuO_x(7)$ and $Gd_{23}Co_{77}(10)/CuO_x(6)$ are both greatly enhanced at low temperature. The nearly constant behavior of $\xi_{DL}^E$ in $Gd_{23}Co_{77}(10)/Pt(5)$ is expected because the intrinsic spin Hall conductivity of Pt is not very sensitive to temperature, consistent with previous measurements of spin torques in RE-TM/Pt bilayers [39]. The OREE in $CuO_x$, on the other hand, is known to increase at low temperature [17,63], which has been attributed to the decrease of phonon-mediated electron hopping between Cu and O orbitals with different symmetry [63]. However, the increase of $\xi_{DL}^E$ is "only" a factor 3 in $Co(2)/CuO_x(7)$, and therefore the increase of the OREE alone cannot explain the factor 12 increase of $|\xi_{DL}^E|$ in $Gd_{23}Co_{77}(10)/CuO_x(6)$, from $-4.4 \times 10^4\ \Omega^{-1}m^{-1}$ at 300 K up to $-5.4 \times 10^5\ \Omega^{-1}m^{-1}$ at 20 K. We suggest that this exceptional increase of the orbital torque in $Gd_{23}Co_{77}(10)/CuO_x(6)$ may be due to additional factors that are specific to ferrimagnets. First, the magnetizations of the Gd and Co sublattices have a strong temperature dependence [45]. Thus, the net spin polarization resulting from the orbital-to-spin conversion at the Gd sites is enhanced at low temperature as the magnetic moments of Gd and Co order along a common axis. Second, as the degree of antiferromagnetic order increases at low temperature, the diffusion length of the orbital and spin currents may increase because the staggered antiferromagnetic exchange field at the Gd and Co sites reduces scattering and spin dephasing [64,65] leading to an increase of the net torque in the relatively thick $Gd_yCo_{100-y}$ layer. This last point is still debated in the literature, as it has been shown that the spin relaxation rates towards the lattice and magnetization compete in establishing a spin torque [66]. Overall, both the orbital-to-spin conversion rate and relaxation rates should be taken into account to achieve a detailed understanding of the orbital torque in ferrimagnets.

Measurements of $\xi_{DL}^E$ in $Gd_{15}Co_{85}(10)/CuO_x(6)$ show that the torque remains largely compensated as a function of temperature, whereas it increases while staying positive in $Gd_{10}Co_{90}(10)/CuO_x(6)$ [Fig. 4(b)]. Thus the sign of the torque is determined by the relative amount of Gd and Co atoms. However, the relative rise of $|\xi_{DL}^E|$ with decreasing temperature, exemplified by the ratio $[\xi_{DL}^E(50\ K) - \xi_{DL}^E(300\ K)]/\xi_{DL}^E(300\ K)$, increases monotonically with Gd content [45]. This increase correlates neither with the change in resistance of the samples nor with $M_s$. Rather, a mean field model of $M_{Gd}$ and $M_{Co}$ as a function of composition and temperature shows that the increase of torque correlates with the degree of magnetic order in both magnetic sublattices [45]. This dependence is reminiscent of the skew scattering contribution of rare-earth impurities to the anomalous Hall conductivity of transition metals, which scales with their magnetic polarization [67]. Although we cannot separate the effects of $M_{Gd}$ and $M_{Co}$ on $\xi_{DL}^E$, our analysis confirms that the orbital-to-spin conversion depends strongly on the degree of magnetic order in the Gd and Co sublattices and their antiferromagnetic coupling.

Remarkably, taking Pt as the standard reference for a spin current source, $\xi_{DL}^E$ of $Gd_{23}Co_{77}(10)/CuO_x(6)$ at 20 K is about a factor 2 larger compared to $Gd_{23}Co_{77}(10)/Pt(5)$. The comparison of the two samples in terms of the injected current density $j = E/\rho$, where $\rho$ is the average resistivity of the bilayers, shows that the torque efficiency normalized to the injected current, i.e., the effective spin-orbital Hall angle $\xi_{DL}^j = \xi_{DL}^E \rho$, changes from -0.02 at 300 K to -0.25 at 20 K in $Gd_{23}Co_{77}(10)/CuO_x(6)$, whereas it remains nearly constant around 0.13 in $Gd_{23}Co_{77}(10)/Pt(5)$ [45]. On the other hand, the resistance of the two samples decreases by 10% and 30% in the same temperature range, respectively. We also note that the composition and temperature dependence of the orbital $\xi_{DL}^j$ differ from those reported for spin torques in ferrimagnets, whereby $\xi_{DL}^j$ is either constant [37,38,68,69] or tends to zero at the compensation point [66].

Last, we point out that $B_{DL}$ and $\xi_{DL}^E$ do not change sign across the magnetic compensation point ($y \approx 27\%$ at room temperature) and magnetic compensation temperature ($\approx 120$ K and 200 K at $y$ = 23% and 24%, respectively). This behavior is consistent with that expected for the injection of angular momentum from an external source, which does not change sign as a function of composition and temperature of GdCo, as do neither the net saturation magnetization nor the spin-orbit coupling of Gd and Co [45].

We conclude that ferrimagnetic alloys offer multiple perspectives to tune the magnitude and sign of the orbital torque in bilayer systems. Our measurements show that the orbital torque changes from positive to negative in Co/CuO$_x$ compared to Gd$_y$Co$_{100-y}$/CuO$_x$ and that $\xi_{DL}^E$ changes monotonically with a negative slope with Gd content in Gd$_y$Co$_{100-y}$. We attribute this behavior to the highly efficient orbital-to-spin conversion introduced by the Gd atoms with negative spin-orbit coupling. The strong enhancement of both $\xi_{DL}^E$ and $\xi_{DL}^j$ observed at low temperature in Gd$_y$Co$_{100-y}$/CuO$_x$ relative to Co/CuO$_x$ and Gd$_y$Co$_{100-y}$/Pt further indicates that the degree of magnetic order in the Gd and Co sublattices plays a significant role in increasing the orbital-to-spin conversion rate. These results highlight the mechanisms that promote orbital-to-spin conversion in a magnetic layer at an atomistic level and call for a more detailed understanding of the conversion process in terms of local versus band structure effects. Our work further demonstrates the competition of different orbital-to-spin conversion processes in RE-TM alloys, providing a versatile method to engineer the orbital torque in ferrimagnets and maximize their efficiency.

**Acknowledgments**

This work was funded by the Swiss National Science Foundation (Grant No. 200020_200465). W.L. acknowledges support from the ETH Zurich Postdoctoral Fellowship programme (21-1 FEL-48). M.-G.K. acknowledges support from the Basic Science Research Program of the National Research Foundation of Korea (Grant No. 2022R1A6A3A03053958).

**FIGURES**

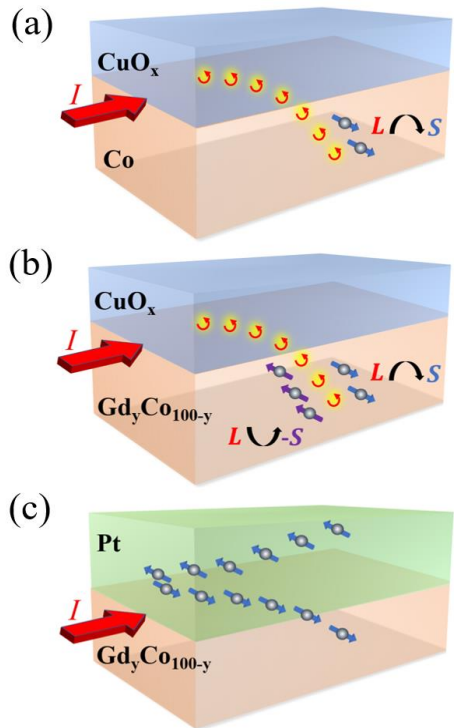

Fig. 1. Schematics of current-induced orbital torques and spin torques. (a) The nonequlibrium orbital moment generated by an electric current in CuO$_x$ due to the OREE ($L$, circular arrows) diffuses into a Co layer, where it is converted into a spin moment ($S$, straight arrows) due to the spin-orbit coupling of Co. (b) In Gd$_y$Co$_{100-y}$, the opposite spin-orbit coupling on the Co and Gd sites results in competing orbital-to-spin conversion. (c) The nonequilibrium spin moment generated by an electric current in Pt due to the spin Hall effect diffuses in Gd$_y$Co$_{100-y}$, where it directly couples to the local magnetization to exert a spin torque.

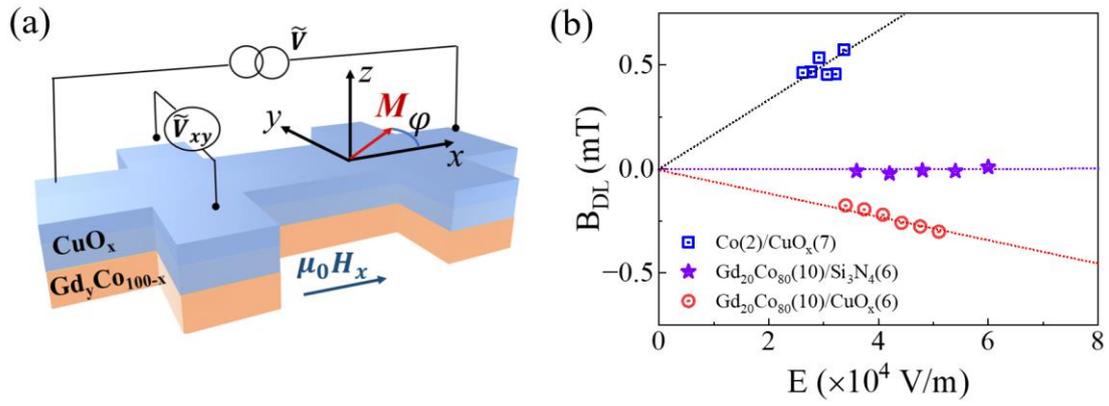

Fig. 2. (a) Schematics of the sample and coordinate system. (b) Effective magnetic field corresponding to the damping-like orbital torque in Co(2)/CuO$_x$(7), Gd$_{20}$Co$_{80}$(10)/CuO$_x$(6), and Gd$_{20}$Co$_{80}$(10) capped by Si$_3$N$_4$(6) as a function of the applied electric field. The dotted lines are linear fits to the data constrained through the origin. The measurements were performed at room temperature.

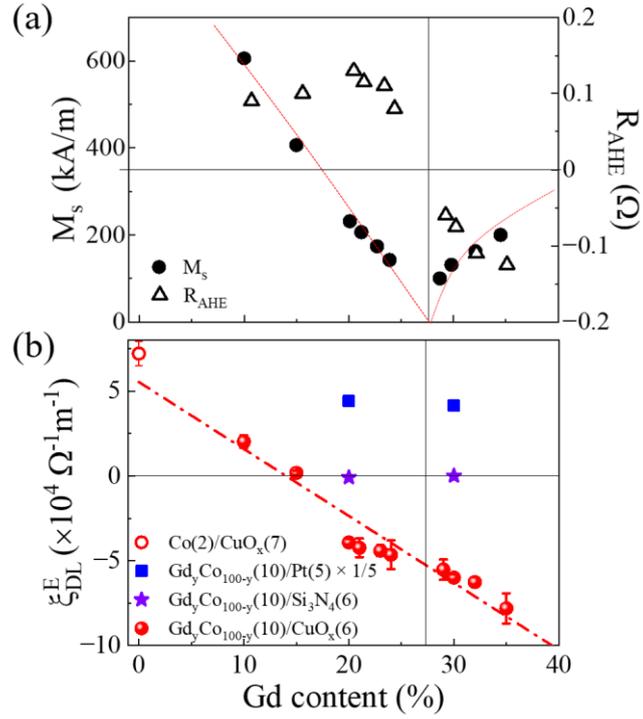

Fig. 3. (a) Saturation magnetization $M_s$ and anomalous Hall resistance $R_{AHE}$ as a function of Gd content in $Gd_yCo_{100-y}(10)/CuO_x(6)$. $M_s$ tends to zero and $R_{AHE}$ changes sign across the magnetic compensation point (vertical solid line). The dashed line is a guide to the eye. (b) Orbital torque efficiency of $Gd_yCo_{100-y}(10)/CuO_x(6)$ vs Gd content (red circles). The dot-dashed line is a linear fit according to Eq. (2). Data for $Co(2)/CuO_x(7)$ (open circle), $Gd_yCo_{100-y}(10)/Pt(5)$ (squares), and $Gd_yCo_{100-y}(10)/Si_3N_4(6)$ (stars) are also shown, all measured at room temperature.

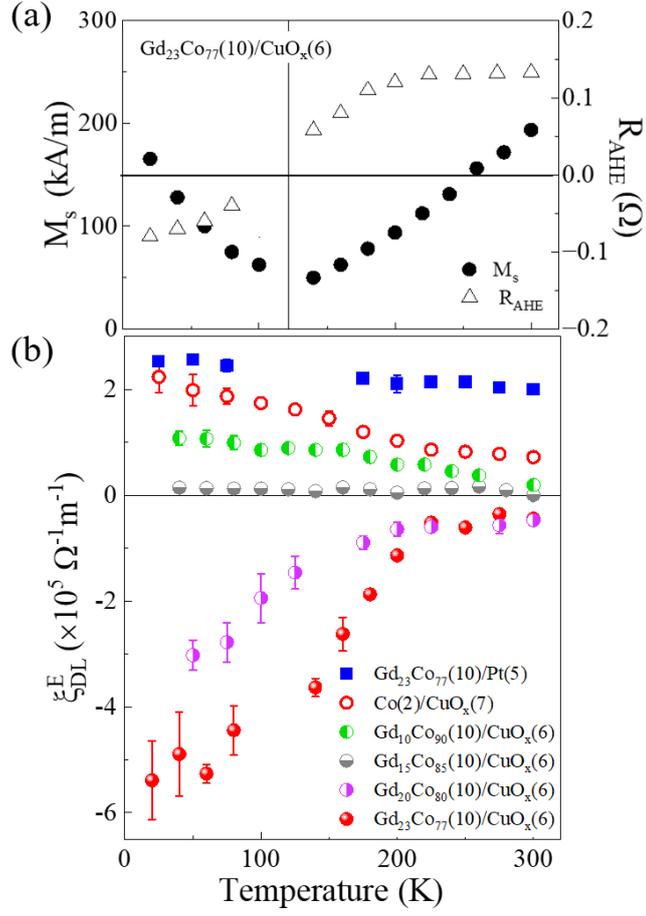

Fig. 4. (a) $M_s$ and $R_{AHE}$ as a function of temperature in Gd$_{23}$Co$_{77}$(10)/CuO$_x$(6). (b) Temperature dependence of the orbital torque efficiency $\xi_{DL}^{E}$ in Co(2)/CuO$_x$(7), Gd$_{23}$Co$_{77}$(10)/Pt(5) and Gd$_y$Co$_{100-y}$(10)/CuO$_x$(6).